\documentclass[aps,prl,twocolumn,showpacs,superscriptaddress]{revtex4-1}

\usepackage{graphics}% Include figure files
\usepackage{graphicx}% Include figure files
\usepackage{dcolumn}% Align table columns on decimal point
\usepackage{bm}% bold math
\usepackage{color}
\usepackage{cancel}
\usepackage{amssymb,amsmath}

\begin{document}

\title{Temporal condensed matter physics in gas-filled photonic crystal fibers}
\author{Mohammed F. Saleh}
\affiliation{Max Planck Institute for the Science of Light, G\"{u}nther-Scharowsky str. 1, 91058 Erlangen, Germany}
\affiliation{School of Engineering and Physical Sciences, Heriot-Watt University, EH14 4AS Edinburgh, UK}
\author{Andrea Armaroli}
\affiliation{Max Planck Institute for the Science of Light, G\"{u}nther-Scharowsky str. 1, 91058 Erlangen, Germany}
\author{Truong X. Tran}
\affiliation{Max Planck Institute for the Science of Light, G\"{u}nther-Scharowsky str. 1, 91058 Erlangen, Germany}
\affiliation{Department of Physics, Le Quy Don University, Vietnam}
\author{Andrea Marini}
\affiliation{Max Planck Institute for the Science of Light, G\"{u}nther-Scharowsky str. 1, 91058 Erlangen, Germany}
\author{Federico Belli}
\affiliation{Max Planck Institute for the Science of Light, G\"{u}nther-Scharowsky str. 1, 91058 Erlangen, Germany}
%\author{John C. Travers}
\author{Amir Abdolvand}
\affiliation{Max Planck Institute for the Science of Light, G\"{u}nther-Scharowsky str. 1, 91058 Erlangen, Germany}
%\author{Philip St.J. Russell}
%\affiliation{Max Planck Institute for the Science of Light, G\"{u}nther-Scharowsky str. 1, 91058 Erlangen, Germany}
%\affiliation{Department of Physics, University of Erlangen-Nuremberg, Germany}
\author{Fabio Biancalana}
\affiliation{Max Planck Institute for the Science of Light, G\"{u}nther-Scharowsky str. 1, 91058 Erlangen, Germany}
\affiliation{School of Engineering and Physical Sciences, Heriot-Watt University, EH14 4AS Edinburgh, UK}
\date{\today}

\begin{abstract}
Raman effect in gases can generate an extremely long-living wave of coherence that can lead to the establishment of an almost perfect periodic variation of the medium refractive index. We show theoretically and numerically that the equations, regulate the pulse propagation in hollow-core photonic crystal fibers filled by Raman-active gas, are exactly identical to a classical problem in quantum condensed matter physics -- but with the role of space and time reversed -- namely an electron in a periodic potential subject to a constant electric field. We are therefore able to infer the existence of Wannier-Stark ladders, Bloch oscillations,  and Zener tunneling, phenomena that are normally associated with condensed matter physics only, now realized with purely optical means in the temporal domain.

\end{abstract}
\maketitle

Hollow-core photonic crystal fibers (HC-PCFs) continue to demonstrate their enormous potential for developing novel photonic devices for different optical applications \cite{Russell03,Russell06}. HC-PCFs with Kagome-style cladding structure have granted unprecedented strong interactions between light and gaseous media over relatively-long propagation distances with low transmission losses and pressure-tunable dispersion in the visible region \cite{Travers11}. Within approximately a decade of the invention of the HC-PCF, exceptional nonlinear phenomena have been demonstrated and predicted in these kind of fibers such as Stokes generation with drastical reduction in the Raman threshold \cite{Benabid02a}, high harmonic generation \cite{Heckl09}, efficient deep-ultraviolet radiation \cite{Joly11}, ionization-induced soliton self-frequency blueshift \cite{Chang11,Hoelzer11b,Saleh11a,Chang13}, strong asymmetrical self-phase modulation, universal modulational instability \cite{Saleh12}, and built-in parity-time symmetry \cite{Saleh14}. 

The Raman effect is one of the earliest and most fundamental effects in nonlinear optics. When a pulse propagates inside a nonlinear medium, it excites optical phonons, leading to a plethora of phenomena, such as the generation of Stokes and anti-Stokes sidebands for long input pulses, and to an intense redshift for short pulses. The Raman effect has been used successfully to enhance the bandwidth of the output spectra during the so-called supercontinuum generation, a dramatic spectral broadening due to the interaction of optical solitons, typically observed in solid-core optical fibers.

Stimulated Raman scattering processes are unique in gases, since they are characterized by a very long molecular coherence relaxation (dephasing) time, of the order of hundreds of picoseconds or more, which should be compared to the short relaxation time of phonon oscillations in silica glass (approx. $32$ fs). Within this long relaxation phase, the medium exhibits a highly non-instantaneous response to pulsed excitations. Nonlinear interactions between optical pulses and Raman-active gases have been suggested and exploited mainly in the synthesis of subfemtosecond pulses using different techniques \cite{Yoshikawa93,Kaplan94,Kawano98,Nazarkin99,Kalosha00}. In addition, continuous down-shift of the frequency of an ultrashort pulse \cite{Korn98}, and optical modulation of a continuous wave laser have been demonstrated due to these interactions \cite{Ihara06}. Recently interaction of an ultrashort pulse with a hydrogen-filled HC-PCF is used by some of the current authors to generate a supercontinuum extending from infrared to vacuum ultraviolet \citep{Belli14} . In the impulsive excitation regime, when the temporal pulse width is shorter than the Raman oscillation period of the gas, a sinusoidal temporal modulation of the medium refractive index lagging the pump has been observed experimentally \cite{Korn98,Nazarkin99,Wittmann00}. This modulation can be detected via launching a delayed weak probe within the dephasing time. 

In this Letter, we analyze the propagation of two temporally separated pulses in HC-PCFs filled with Raman-active gases. By investigating the effect of the Raman polarization induced by a pump pulse in the impulsive excitation regime, we demonstrate a perfect analogy between the spatiotemporal dynamics of a delayed probe and several phenomena observed in condensed matter physics, such as the Wannier-Stark ladder \cite{Wannier60}, Bloch oscillations \cite{Bloch28} and Zener tunneling \cite{Zener34}.

\paragraph*{Governing equations for Raman media ---} Let us consider the propagation of an ultrashort pulse in a HC-PCF filled with a Raman-active gas. The medium response is described by its total polarization, which is a sum of the linear, Kerr, and Raman polarizations. Ionization-induced plasma generation is neglected, as we assume that the pulse intensity is smaller than the gas ionization threshold. Assuming that only one Raman mode is excited (either the vibrational or rotational one), the dynamics of the Raman polarization (also called {\em coherence}) $ P_{R} $ can be determined by solving the Bloch equations for a two-level system \cite{Butylkin89,Kalosha00}:
\begin{equation}
\begin{array}{l}
\partial_{t} w + \dfrac{w+1}{T_{1}} =\dfrac{i\alpha_{12}}{\hbar}\left(\rho_{12}-\rho_{12}^{*} \right)E^{2}, \\ 
\left[ \partial_{t}  + \dfrac{1}{T_{2}}-i\omega_{\rm R}\right]\rho_{12} =\dfrac{i}{2\hbar}\left[\alpha_{12} w + \left(\alpha_{11}-\alpha_{22} \right)\rho_{12} \right]E^{2},
\end{array}
\end{equation}
where $ w=\rho_{22}-\rho_{11} $ is the population inversion between the excited and ground states, $ t $ is the time variable, $ \alpha_{ij} $ and $ \rho_{ij} $ are the elements of the $ 2\times 2 $ polarizability and density matrices, respectively, $ E $ is the real electric field, $ \omega_{\rm R} $ is the Raman frequency of the transition, $ N_{0} $ is the molecular number density, $ T_{1} $ and $ T_{2} $ are the population and polarization relaxation times, respectively, and $ \hbar $ is the reduced Planck's constant. Finally we have that $ P_{\rm R}\approx \left[ \alpha_{12}  \left(\rho_{12}+\rho_{12}^{*} \right)+\alpha_{11}\rho_{11}+\alpha_{22}\rho_{22}\right]  N_{0}E$, assuming that initially all the molecules are in the ground state. In the slowly varying envelope approximation (SVEA) \cite{Agrawal07}, the electric field can be expressed as $ E= \frac{1}{2}\left[ A\left(z,t \right)  \exp\left(i\beta_{0} z-i\omega_{0}t \right) +c.c. \right] $, where $ \omega_{0} $ is the pulse central frequency, $ \beta_{0}$ is the propagation constant calculated at $\omega_{0}$, $ z $ is the longitudinal coordinate along the fiber, $ A $ is the complex envelope, and $ c.c. $ denotes the complex conjugate. We first introduce the following new variables: $ \xi=z/z_{0} $, $ \tau=t/t_{0} $, $ \psi=A/A_{0} $, $ z_{0}=t_{0}^{2}/\left|\beta_{2}\left( \omega_{0}\right)\right|  $, and $ A_{0}^{2}=1/\left(\gamma z_{0} \right) $, where $ \gamma $ is the nonlinear Kerr coefficient in the unit of W$^{-1}$m$ ^{-1} $, $ \beta_{2} $ is the second-order dispersion coefficient, and $ t_{0} $ is the pulse duration. By using the SVEA, one can derive the following set of coupled equations:
\begin{equation}
\begin{array}{l}
\left[ i\partial_{\xi}+\hat{D}(i\partial_{\tau})+|\psi|^{2} +\eta\,\mathrm{Re}\left(\rho_{12} \right)\right] \psi  =0,  \\
\partial_{\tau} w + \dfrac{\left( w+1\right)t_{0} }{T_{1}} =-4\, \mu\, w\, \mathrm{Im}\left(\rho_{12} \right)\, \left|\psi\right|^{2},\\ 
\left[ \partial_{\tau}  + \dfrac{t_{0}}{T_{2}}-i\delta\right]\rho_{12} =i\mu\, w\, \left|\psi\right|^{2},
\end{array}
\label{eq1}
\end{equation}
where $\hat{D}(i\partial_{\tau})=\sum_{m\geq 1}\beta_{m}(i\partial_{\tau})^{m}t_{0}^{m}/m!$ is the full dispersion operator, $\beta_{m}$ is the $m$th order dispersion coefficient calculated at $\omega_{0}$, $ \eta=z_{0}/z_{1} $, $ z_{1}= c\,\epsilon_{0}/\left(\alpha_{12}\,N_{0}\,\omega_{0} \right)  $ is the nonlinear Raman length, $ \mu=P_{0}/P_{1} $, $ P_{1}=2\hbar\, c\,\epsilon_{0} A_{\mathrm{eff}} /\left(\alpha_{12}\,t_{0} \right) $, $ \delta= \omega_{\rm R}t_{0}$, $ c $ is the speed of light, $ \epsilon_{0} $ is the vacuum permittivity, $ A_{\mathrm{eff}} $ is the effective area of the fundamental mode, and Re, Im represent the real and imaginary parts. In the derivation of Eqs. (\ref{eq1}), we have assumed that $ \left|w\right|\gg \left|\rho_{12}\right|$, and the population inversion is weak, i.e. $ \rho_{11} \approx 1$, and $ \rho_{22}\approx 0 $. These assumptions are physically very realistic in gaseous Raman media.

\begin{figure}
\includegraphics[width=8.6cm]{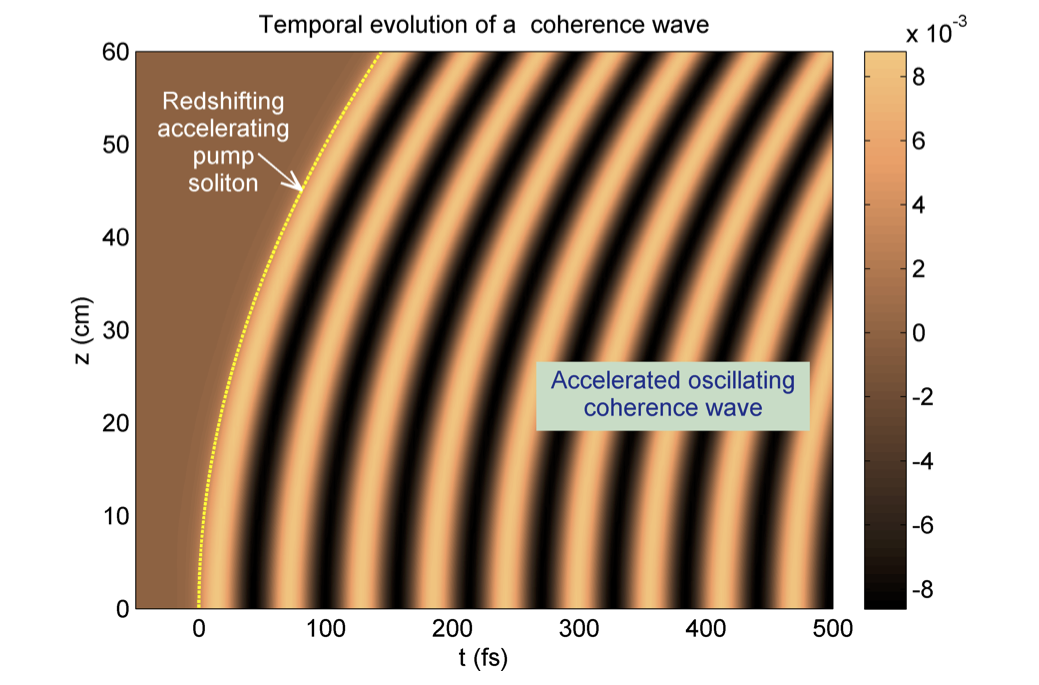}
\caption{(Color online). Temporal evolution of an accelerated oscillating Raman polarization with period $ \Lambda=56.7 $ fs by a propagating fundamental soliton with amplitude $V_{1}=1.33$, central wavelength $1064$ nm, and full width at half maximum (FWHM) $15$ fs in a H$ _{2} $-filled HC-PCF that has a flat-to-flat core diameter 18 $ \mu $m, gas pressure $7$ bar, and rotational Raman frequency $ \omega_{\rm R} = 17.6 $ THz. The dashed yellow line represents the temporal evolution of the soliton that excites the coherence wave. The simulation parameters are $ \gamma=7.07$ MW$ ^{-1} $m$ ^{-1} $, $ \beta_{2}=-3425.5 $ fs$ ^{2} $/m, $ A_{\mathrm{eff}}=134\,\mu $m$ ^{2} $, $ \alpha_{12}=0.8 \times 10^{-41} $ C m$^{2}$/V, and $ t_{0}= $  11.34 fs. All calculations in this Letter are based on these values.
\label{Fig0}}
\end{figure}

\paragraph*{Raman response function ---}
For femtosecond pulses, the relaxation times of the population inversion ($T_{1}$) and the coherence ($T_{2}$) can be safely neglected, since they are of the order of hundreds of picoseconds or more. This amazingly long relaxation times are a crucial and quite unique feature of gaseous Raman systems, which we use to the full in the present Letter. We have also found that the population inversion is almost unchanged from its initial value for pulses with energies in the order of few $ \mu $J, i.e. $ w\left(\tau \right)\approx w\left(-\infty \right)=-1 $. Hence, the set of the governing equations Eq. (\ref{eq1}) can be reduced to a single generalized nonlinear Schr\"{o}dinger equation
\begin{equation}
i\partial_{\xi}\psi+i\frac{\beta_{1}z_{0}}{t_{0}}\partial_{\tau}\psi+\frac{1}{2}\partial_{\tau}^{2}\psi+|\psi|^{2}\psi   +R\left(\tau \right)\psi=0,
\label{eq2}
\end{equation}
where pumping in the deep anomalous dispersion regime ($\beta_{2}<0$) is assumed, higher-order dispersion coefficients $ \beta_{m>2} $ are neglected, $ R\left(\tau \right)=\kappa \int_{-\infty}^{\tau}\sin\left[ \delta\left( \tau-\tau'\right) \right]  \left|\psi\left( \tau'\right) \right|^{2} d\tau'$ is the resulting Raman convolution, and $ \kappa=\eta\mu $. For ultrashort pulses with duration $t_{0}\ll1/\omega_{\rm R}$, $\sin\left[ \delta\left( \tau-\tau'\right) \right] $ can be expanded around the temporal location of the pulse peak by using a Taylor expansion. For instance, a fundamental soliton with amplitude $ V $ and centered at $ \tau=0 $ will induce a Raman contribution in the form of $ R\left(\tau \right)\approx \kappa V\sin\left( \delta\tau\right) \left [ 1+ \,\mathrm{tanh} \left(V\tau\right)  \right] $ at the zeroth-order approximation. Hence, this soliton will induce a retarded sinusoidal Raman polarization that can impact the dynamics of the other trailing probe pulse lagging behind the pump pulse, see Fig. \ref{Fig0}. On the other hand, for $t_{0}\gg1/\omega_{\rm R}$, $R\left(\tau \right)\approx \gamma_{R}\left|\psi\left( \tau\right) \right|^{2}$ with $ \gamma_{R}= \kappa/\delta$, i.e the Raman nonlinearity can be considered instantaneous. So, the Raman contribution would induce an effective Kerr nonlinearity that is significant, and, in many cases, can compete directly with the intrinsic Kerr nonlinearity of the gas.

\paragraph*{Pump solution ---} We focus on two different pulses that are not overlapped in time, and separated by a delay $\ll T_{1}, T_{2}  $ in the deep anomalous dispersion regime. The leading pulse is an ultrashort strong `pump' pulse $ \psi_{1}$ with $t_{0}\ll1/\omega_{\rm R}$, while the trailing pulse is a weak `probe' pulse $ \psi_{2}$ with negligible nonlinearity. In this case, Eq. (\ref{eq2}) can be used to determine the pump solution. For weak Raman nonlinearity, the solution of Eq. (\ref{eq2}) can be assumed to be a fundamental soliton that is perturbed by the Raman polarization, i.e. $ \psi_{1}\left( \xi,\tau\right)=V_{1}\,\mathrm{sech} \left[V_{1} \left(\tau-u_{1}\xi-\bar{\tau}_{1}\left(\xi \right) \right) \right] \exp\left[-i\Omega_{1}\left(\xi \right)\left(\tau-u_{1}\xi\right)\right] $ where $u_{1}=\beta_{11}z_{0}/t_{0}$,  $\beta_{11}$ is the first-order dispersion coefficient of the pump, $V_{1} $, $ \Omega_{1} $, and $\bar{\tau}_{1}  $  are the soliton amplitude, central frequency, and position of its peak, respectively. In the following, we will assume launching a pump soliton with these parameters, and with $\bar{\tau}_{1}\left(0 \right)=0$. Using the variational perturbation method \cite{Agrawal07}, this soliton is found to be linearly redshifting in the frequency domain with rate $ g_{1} =\frac{1}{2}\kappa\pi\delta^{2} \mathrm{csch}\left( \pi\delta/2V_{1}\right)$, and decelerating in the time domain, i.e. $ \Omega_{1} = - g_{1}\, \xi,$ and $ \bar{\tau}_{1} = g_{1}\, \xi^{2}/2 $.

\paragraph*{Governing equation for the probe ---}  When a second weak probe pulse is sent after the leading pump soliton described in the previous section, the probe evolution is ruled by the equation
\begin{equation}
 i\partial_{\xi}\psi_{2}+i u_{2}\partial_{\tau}\psi_{2}+\dfrac{1}{2m}\partial_{\tau}^{2}\psi_{2} +2\kappa V_{1}\sin\left( \delta\tilde{\tau}\right) \psi_{2}  = 0 ,     
\label{eq3}
\end{equation}
where $ u_{2}= \beta_{12} z_{0}/t_{0}  $, $ m=\left|\beta_{21}\right|/\left|\beta_{22}\right| $, $ \beta_{1j} $ and $ \beta_{2j} $ are the first and the second order dispersion coefficients of the $ j^{\mathrm{th}} $ pulse with $ j=1,2 $. Going to the reference frame of the leading decelerating soliton, $ \tilde{\tau}=\tau-u_{1}\xi-g_{1}\xi^{2}/2 $, and applying a generalized form of the Gagnon-B\'{e}langer phase transformation \cite{Gagnon90} $ \psi_{2}\left( \xi,\tilde{\tau}\right) =\phi\left(\xi,\tilde{\tau} \right) \exp\left[i\tilde{\tau}\left(g_{1}\xi +u_{1}-u_{2} \right) +i \left(g_{1}\xi+u_{1}-u_{2} \right)^{3} /6g_{1}\right]  $, Eq. (\ref{eq3}) becomes
\begin{equation}
i\partial_{\xi}\phi=-\frac{1}{2m}\partial_{\tilde{\tau}}^{2}\phi  + \left[-2\kappa V_{1}\sin \left(\delta\tilde{\tau} \right)+g_{1}\tilde{\tau}\right] \phi.
\label{eq4}
\end{equation}
This equation is the {\em exact analogue} of the time-dependent Schr\"{o}dinger equation of an electron in a periodic crystal in the presence of an external electric field. In Eq. (\ref{eq4}) time and space are swapped with respect to the condensed matter physics system, as usual in optics, and we deal with a spatial-dependent Schr\"{o}dinger equation of a single particle `probe' with mass $ m $ in a \textit{temporal crystal} with a periodic potential $ U= -2\kappa V_{1}\sin \left(\delta\tilde{\tau} \right) $ in the presence of a constant force $ -g_{1} $ in the positive-delay direction. The leading soliton excites a sinusoidal Raman oscillation that forms a periodic structure in the reference frame of the soliton. Due to soliton acceleration induced by the strong spectral redshift, a constant force is applied on this structure. Substituting $ \phi\left(\xi,\tilde{\tau} \right) = f\left(\tilde{\tau}\right) \exp\left(  iq\xi\right)  $, Eq. (\ref{eq4}) becomes an eigenvalue problem with eigenvectors $ f $, and eigenvalues $ -q $. The modes of this equation are the Wannier functions \cite{Wannier60} that can exhibit Bloch oscillations \cite{Bloch28}, intrawell oscillations \cite{Bouchard95}, and Zener tunneling \cite{Zener34} due to the applied force. Note that here we cannot use the so-called Houston functions (see Ref. \cite{Iafrate98}) for solving Eq. (\ref{eq4}), since these functions are valid only for small external electric fields - which is not the case in Eq. (\ref{eq4}), since the `artificial electric field' $g_{1}$ is not independent from the depth of the periodic potential $\kappa$. We must therefore calculate numerically the full solution of Eq. (\ref{eq4}).

\begin{figure}
\includegraphics[width=8.6cm]{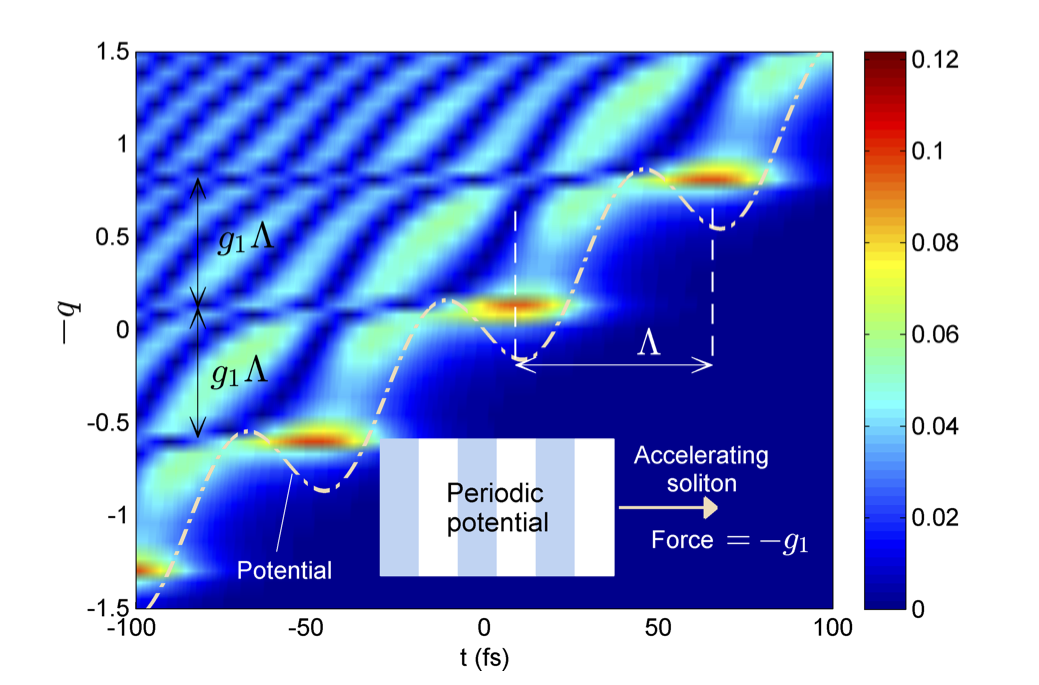}
\caption{(Color online). A portion of the absolute eigenstates of a Raman-induced temporal periodic crystals with a lattice constant $ \Lambda=56.7 $ fs in the presence of a force with magnitude $ g_{1}=0.1408 $ in the positive-delay direction. The vertical axis represents the corresponding eigenvalues $ -q $. The dotted-dashed line is the potential. 
\label{Fig1}}
\end{figure}

\paragraph*{Wannier-Stark ladder ---}
As a practical example, let us consider the propagation of an ultrashort soliton in a H$ _{2} $-filled HC-PCF. Exciting the rotational Raman shift frequency in the fiber via this soliton will induce a long-lived trailing {\em temporal periodic crystal} with a lattice constant $ \Lambda= 56.7$ fs, corresponding to the time required by the H$  _{2}$ molecule to complete one cycle of rotation. In the absence of the applied force, the solutions are the Bloch modes, while in the presence of the applied force, the periodic potential is tilted, and the eigenstates of the system are the Wannier functions portrayed as a  2D color plot in Fig. \ref{Fig1}, where the horizontal axis is the time and the vertical axis is the corresponding eigenvalue. These functions are modified Airy beams that have strong or weak oscillating decaying tails. After an eigenvalue step $ g_{1}\Lambda  $, the eigenstates are repeated, but shifted by $ \Lambda $, forming the Wannier-Stark ladder. As shown, each potential minimum can allow a single localized state with weak tails. A large number of delocalised modes with long and strong tails exist between the localised states. 

\begin{figure}
\includegraphics[width=8.6cm]{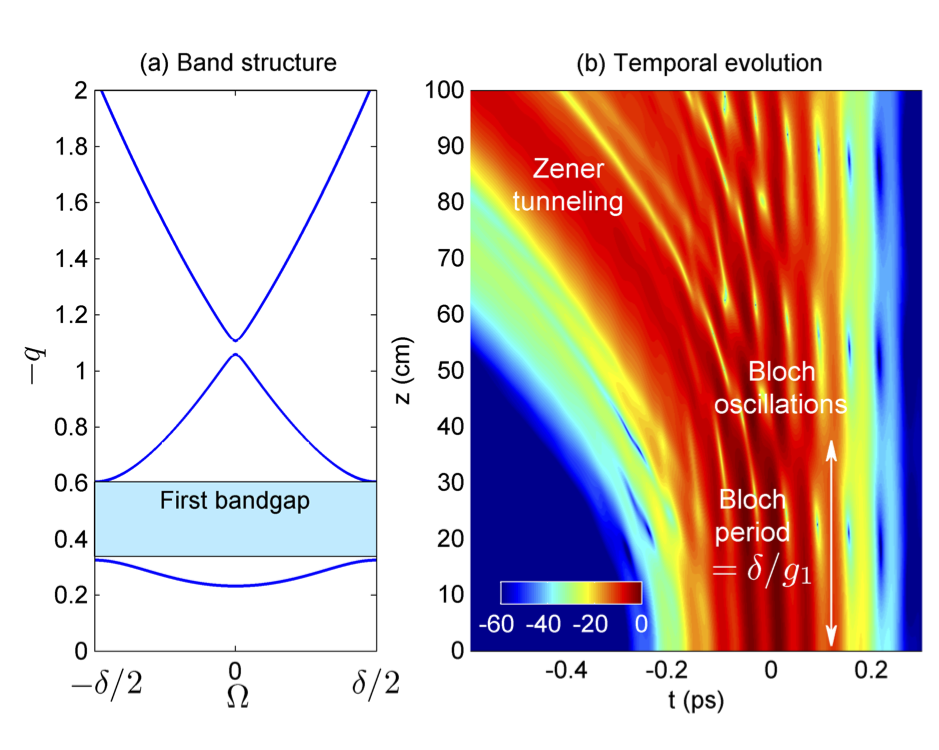}
\caption{(Color online). (a) Bandstructure of the temporal crystal induced by the leading ultrashort soliton propagating in the H$ _{2} $-filled HC-PCF with $ m=1 $. (b) Temporal evolution of a weak probe in the accelerated periodic temporal crystal. The probe initial temporal profile is a Gaussian pulse with FWHM 133.6 fs superimposed on the first Bloch mode of the periodic crystal in the absence of the applied force. The contour plot is given in a logarithmic scale and truncated at -60 dB.
\label{Fig2}}
\end{figure}

\paragraph*{Bloch oscillations and Zener tunneling ---}
An arbitrary weak probe following the soliton will be decomposed into the Wannier modes of the periodic temporal crystal. Due to beating between similar eigenstates in different potential wells, Bloch oscillations arise with a period $ \delta/g_{1} $, while beating between different eigenstates in the same potential minimum can result in intrawell oscillations.  In our case we did not observe in the simulations the latter kind of beating, since only a single eigenstate is allowed within each well. Interference between modes lying between different wells are responsible for Zener tunneling that allows transitions between different wells (or bands). In the absence of the applied force ($g_1=0$), the band structure of the periodic medium can be constructed by plotting the propagation constants of the Bloch modes over the first Brillouin zone $ \left[-\delta/2,\delta/2\right]  $, as shown in Fig. \ref{Fig2}(a). Zener tunneling occurs when a particle transits from the lowest band to the next-higher band. The evolution of a delayed probe in the form of the first Bloch mode inside a H$ _{2} $-filled HC-PCF under the influence of the pump-induced temporal periodic crystal, is depicted in Fig. \ref{Fig2}(b). Portions of the probe are localized in different temporal wells. We show in Fig. \ref{Fig1} Bloch oscillations with period 34.7 cm, which correspond to the beating between modes localized in adjacent wells. After each half of this period, an accelerated radiation to the left due to Zener tunneling is also emitted. The Zener tunneling is dominant, and the Bloch oscillations are weak, because the potential wells are relatively far from each other. The overlapping between the localized modes are small, consistent with the shallowness of the first band in the periodic limit. 

\paragraph*{Conclusions ---}
From the Maxwell-Bloch equations we have derived a model based on SVEA that is convenient for investigating pulse propagation in HC-PCFs filled by Raman-active gases.We have specialized the model to study the propagation of two pulses that do not temporally overlap and are separated by a time delay smaller than the Raman polarization dephasing time ($\sim100$ ps). The leading pulse is an ultrashort strong soliton acting as a pump that experiences linear redshift and deceleration due to Raman nonlinearity. Closed forms of the pump temporal and spectral dynamics have been obtained. The induced-Raman excitation creates a temporal crystal susceptible to a constant force due to soliton acceleration. The trailing pulse is assumed to be a weak probe. In this case, the problem is reduced to the motion of a particle in a periodic crystal subject to an external force. Phenomena such as Wannier-stark ladder, Bloch oscillations, and Zener tunneling have been demonstrated by simulations.  Our results open new research pathways at the border of nonlinear photonics and condensed-matter physics that will bear fruits not only for fundamental science but also for the conception of new devices.

F. B. and M. S. would like to acknowledge several useful discussions with Prof. Philip St.J. Russell and Dr. John Travers at MPL Erlangen.


\begin{thebibliography}{29}%
\makeatletter
\providecommand \@ifxundefined [1]{%
 \@ifx{#1\undefined}
}%
\providecommand \@ifnum [1]{%
 \ifnum #1\expandafter \@firstoftwo
 \else \expandafter \@secondoftwo
 \fi
}%
\providecommand \@ifx [1]{%
 \ifx #1\expandafter \@firstoftwo
 \else \expandafter \@secondoftwo
 \fi
}%
\providecommand \natexlab [1]{#1}%
\providecommand \enquote  [1]{``#1''}%
\providecommand \bibnamefont  [1]{#1}%
\providecommand \bibfnamefont [1]{#1}%
\providecommand \citenamefont [1]{#1}%
\providecommand \href@noop [0]{\@secondoftwo}%
\providecommand \href [0]{\begingroup \@sanitize@url \@href}%
\providecommand \@href[1]{\@@startlink{#1}\@@href}%
\providecommand \@@href[1]{\endgroup#1\@@endlink}%
\providecommand \@sanitize@url [0]{\catcode `\\12\catcode `\$12\catcode
  `\&12\catcode `\#12\catcode `\^12\catcode `\_12\catcode `\%12\relax}%
\providecommand \@@startlink[1]{}%
\providecommand \@@endlink[0]{}%
\providecommand \url  [0]{\begingroup\@sanitize@url \@url }%
\providecommand \@url [1]{\endgroup\@href {#1}{\urlprefix }}%
\providecommand \urlprefix  [0]{URL }%
\providecommand \Eprint [0]{\href }%
\providecommand \doibase [0]{http://dx.doi.org/}%
\providecommand \selectlanguage [0]{\@gobble}%
\providecommand \bibinfo  [0]{\@secondoftwo}%
\providecommand \bibfield  [0]{\@secondoftwo}%
\providecommand \translation [1]{[#1]}%
\providecommand \BibitemOpen [0]{}%
\providecommand \bibitemStop [0]{}%
\providecommand \bibitemNoStop [0]{.\EOS\space}%
\providecommand \EOS [0]{\spacefactor3000\relax}%
\providecommand \BibitemShut  [1]{\csname bibitem#1\endcsname}%
\let\auto@bib@innerbib\@empty
%</preamble>
\bibitem [{\citenamefont {{P.~{St.J}.~Russell}}(2003)}]{Russell03}%
  \BibitemOpen
  \bibfield  {author} {\bibinfo {author} {\bibnamefont
  {{P.~{St.J}.~Russell}}},\ }\href@noop {} {\bibfield  {journal} {\bibinfo
  {journal} {Science}\ }\textbf {\bibinfo {volume} {299}},\ \bibinfo {pages}
  {358} (\bibinfo {year} {2003})}\BibitemShut {NoStop}%
\bibitem [{\citenamefont {{P.~{St.J}.~Russell}}(2006)}]{Russell06}%
  \BibitemOpen
  \bibfield  {author} {\bibinfo {author} {\bibnamefont
  {{P.~{St.J}.~Russell}}},\ }\href@noop {} {\bibfield  {journal} {\bibinfo
  {journal} {J. Light. Technol.}\ }\textbf {\bibinfo {volume} {24}},\ \bibinfo
  {pages} {4729} (\bibinfo {year} {2006})}\BibitemShut {NoStop}%
\bibitem [{\citenamefont {Travers}\ \emph {et~al.}(2011)\citenamefont
  {Travers}, \citenamefont {Chang}, \citenamefont {Nold}, \citenamefont
  {Joly},\ and\ \citenamefont {{P.~{St.J}.~Russell}}}]{Travers11}%
  \BibitemOpen
  \bibfield  {author} {\bibinfo {author} {\bibfnamefont {J.~C.}\ \bibnamefont
  {Travers}}, \bibinfo {author} {\bibfnamefont {W.}~\bibnamefont {Chang}},
  \bibinfo {author} {\bibfnamefont {J.}~\bibnamefont {Nold}}, \bibinfo {author}
  {\bibfnamefont {N.~Y.}\ \bibnamefont {Joly}}, \ and\ \bibinfo {author}
  {\bibnamefont {{P.~{St.J}.~Russell}}},\ }\href@noop {} {\bibfield  {journal}
  {\bibinfo  {journal} {J. Opt. Soc. Am. B}\ }\textbf {\bibinfo {volume}
  {28}},\ \bibinfo {pages} {A11} (\bibinfo {year} {2011})}\BibitemShut
  {NoStop}%
\bibitem [{\citenamefont {Benabid}\ \emph {et~al.}(2002)\citenamefont
  {Benabid}, \citenamefont {Knight}, \citenamefont {Antonopoulos},\ and\
  \citenamefont {{P.~{St.J}.~Russell}}}]{Benabid02a}%
  \BibitemOpen
  \bibfield  {author} {\bibinfo {author} {\bibfnamefont {F.}~\bibnamefont
  {Benabid}}, \bibinfo {author} {\bibfnamefont {J.~C.}\ \bibnamefont {Knight}},
  \bibinfo {author} {\bibfnamefont {G.}~\bibnamefont {Antonopoulos}}, \ and\
  \bibinfo {author} {\bibnamefont {{P.~{St.J}.~Russell}}},\ }\href@noop {}
  {\bibfield  {journal} {\bibinfo  {journal} {Science}\ }\textbf {\bibinfo
  {volume} {298}},\ \bibinfo {pages} {399} (\bibinfo {year}
  {2002})}\BibitemShut {NoStop}%
\bibitem [{\citenamefont {Heckl}\ \emph {et~al.}(2009)\citenamefont {Heckl},
  \citenamefont {Baer}, \citenamefont {Kr\"{a}nkel}, \citenamefont {Marchese},
  \citenamefont {Schapper}, \citenamefont {Holler}, \citenamefont
  {S\"{u}dmeyer}, \citenamefont {Robinson}, \citenamefont {Tisch},
  \citenamefont {Couny}, \citenamefont {Light}, \citenamefont {Benabid},\ and\
  \citenamefont {Keller}}]{Heckl09}%
  \BibitemOpen
  \bibfield  {author} {\bibinfo {author} {\bibfnamefont {O.~H.}\ \bibnamefont
  {Heckl}}, \bibinfo {author} {\bibfnamefont {C.~R.~E.}\ \bibnamefont {Baer}},
  \bibinfo {author} {\bibfnamefont {C.}~\bibnamefont {Kr\"{a}nkel}}, \bibinfo
  {author} {\bibfnamefont {S.~V.}\ \bibnamefont {Marchese}}, \bibinfo {author}
  {\bibfnamefont {F.}~\bibnamefont {Schapper}}, \bibinfo {author}
  {\bibfnamefont {M.}~\bibnamefont {Holler}}, \bibinfo {author} {\bibfnamefont
  {T.}~\bibnamefont {S\"{u}dmeyer}}, \bibinfo {author} {\bibfnamefont {J.~S.}\
  \bibnamefont {Robinson}}, \bibinfo {author} {\bibfnamefont {J.~W.~G.}\
  \bibnamefont {Tisch}}, \bibinfo {author} {\bibfnamefont {F.}~\bibnamefont
  {Couny}}, \bibinfo {author} {\bibfnamefont {P.}~\bibnamefont {Light}},
  \bibinfo {author} {\bibfnamefont {F.}~\bibnamefont {Benabid}}, \ and\
  \bibinfo {author} {\bibfnamefont {U.}~\bibnamefont {Keller}},\ }\href@noop {}
  {\bibfield  {journal} {\bibinfo  {journal} {Appl. Phys. B}\ }\textbf
  {\bibinfo {volume} {97}},\ \bibinfo {pages} {369} (\bibinfo {year}
  {2009})}\BibitemShut {NoStop}%
\bibitem [{\citenamefont {Joly}\ \emph {et~al.}(2011)\citenamefont {Joly},
  \citenamefont {Nold}, \citenamefont {Chang}, \citenamefont {H\"{o}lzer},
  \citenamefont {Nazarkin}, \citenamefont {Wong}, \citenamefont {Biancalana},\
  and\ \citenamefont {{P.~{St.J}.~ Russell}}}]{Joly11}%
  \BibitemOpen
  \bibfield  {author} {\bibinfo {author} {\bibfnamefont {N.~Y.}\ \bibnamefont
  {Joly}}, \bibinfo {author} {\bibfnamefont {J.}~\bibnamefont {Nold}}, \bibinfo
  {author} {\bibfnamefont {W.}~\bibnamefont {Chang}}, \bibinfo {author}
  {\bibfnamefont {P.}~\bibnamefont {H\"{o}lzer}}, \bibinfo {author}
  {\bibfnamefont {A.}~\bibnamefont {Nazarkin}}, \bibinfo {author}
  {\bibfnamefont {G.~K.~L.}\ \bibnamefont {Wong}}, \bibinfo {author}
  {\bibfnamefont {F.}~\bibnamefont {Biancalana}}, \ and\ \bibinfo {author}
  {\bibnamefont {{P.~{St.J}.~ Russell}}},\ }\href@noop {} {\bibfield  {journal}
  {\bibinfo  {journal} {Phys. Rev. Lett.}\ }\textbf {\bibinfo {volume} {106}},\
  \bibinfo {pages} {203901} (\bibinfo {year} {2011})}\BibitemShut {NoStop}%
\bibitem [{\citenamefont {Chang}\ \emph {et~al.}(2011)\citenamefont {Chang},
  \citenamefont {Nazarkin}, \citenamefont {Travers}, \citenamefont {Nold},
  \citenamefont {H\"{o}lzer}, \citenamefont {Joly},\ and\ \citenamefont
  {{P.~{St.J}.~Russell}}}]{Chang11}%
  \BibitemOpen
  \bibfield  {author} {\bibinfo {author} {\bibfnamefont {W.}~\bibnamefont
  {Chang}}, \bibinfo {author} {\bibfnamefont {A.}~\bibnamefont {Nazarkin}},
  \bibinfo {author} {\bibfnamefont {J.~C.}\ \bibnamefont {Travers}}, \bibinfo
  {author} {\bibfnamefont {J.}~\bibnamefont {Nold}}, \bibinfo {author}
  {\bibfnamefont {P.}~\bibnamefont {H\"{o}lzer}}, \bibinfo {author}
  {\bibfnamefont {N.~Y.}\ \bibnamefont {Joly}}, \ and\ \bibinfo {author}
  {\bibnamefont {{P.~{St.J}.~Russell}}},\ }\href@noop {} {\bibfield  {journal}
  {\bibinfo  {journal} {Opt. Express}\ }\textbf {\bibinfo {volume} {19}},\
  \bibinfo {pages} {21018} (\bibinfo {year} {2011})}\BibitemShut {NoStop}%
\bibitem [{\citenamefont {H\"{o}lzer}\ \emph {et~al.}(2011)\citenamefont
  {H\"{o}lzer}, \citenamefont {Chang}, \citenamefont {Travers}, \citenamefont
  {Nazarkin}, \citenamefont {Nold}, \citenamefont {Joly}, \citenamefont
  {Saleh}, \citenamefont {Biancalana},\ and\ \citenamefont
  {{P.~{St.J}.~Russell}}}]{Hoelzer11b}%
  \BibitemOpen
  \bibfield  {author} {\bibinfo {author} {\bibfnamefont {P.}~\bibnamefont
  {H\"{o}lzer}}, \bibinfo {author} {\bibfnamefont {W.}~\bibnamefont {Chang}},
  \bibinfo {author} {\bibfnamefont {J.~C.}\ \bibnamefont {Travers}}, \bibinfo
  {author} {\bibfnamefont {A.}~\bibnamefont {Nazarkin}}, \bibinfo {author}
  {\bibfnamefont {J.}~\bibnamefont {Nold}}, \bibinfo {author} {\bibfnamefont
  {N.~Y.}\ \bibnamefont {Joly}}, \bibinfo {author} {\bibfnamefont {M.~F.}\
  \bibnamefont {Saleh}}, \bibinfo {author} {\bibfnamefont {F.}~\bibnamefont
  {Biancalana}}, \ and\ \bibinfo {author} {\bibnamefont
  {{P.~{St.J}.~Russell}}},\ }\href@noop {} {\bibfield  {journal} {\bibinfo
  {journal} {Phys. Rev. Lett.}\ }\textbf {\bibinfo {volume} {107}},\ \bibinfo
  {pages} {203901} (\bibinfo {year} {2011})}\BibitemShut {NoStop}%
\bibitem [{\citenamefont {Saleh}\ \emph {et~al.}(2011)\citenamefont {Saleh},
  \citenamefont {Chang}, \citenamefont {H\"olzer}, \citenamefont {Nazarkin},
  \citenamefont {Travers}, \citenamefont {Joly}, \citenamefont
  {{P.~{St.J}.~Russell}},\ and\ \citenamefont {Biancalana}}]{Saleh11a}%
  \BibitemOpen
  \bibfield  {author} {\bibinfo {author} {\bibfnamefont {M.~F.}\ \bibnamefont
  {Saleh}}, \bibinfo {author} {\bibfnamefont {W.}~\bibnamefont {Chang}},
  \bibinfo {author} {\bibfnamefont {P.}~\bibnamefont {H\"olzer}}, \bibinfo
  {author} {\bibfnamefont {A.}~\bibnamefont {Nazarkin}}, \bibinfo {author}
  {\bibfnamefont {J.~C.}\ \bibnamefont {Travers}}, \bibinfo {author}
  {\bibfnamefont {N.~Y.}\ \bibnamefont {Joly}}, \bibinfo {author} {\bibnamefont
  {{P.~{St.J}.~Russell}}}, \ and\ \bibinfo {author} {\bibfnamefont
  {F.}~\bibnamefont {Biancalana}},\ }\href@noop {} {\bibfield  {journal}
  {\bibinfo  {journal} {Phys. Rev. Lett.}\ }\textbf {\bibinfo {volume} {107}},\
  \bibinfo {pages} {203902} (\bibinfo {year} {2011})}\BibitemShut {NoStop}%
\bibitem [{\citenamefont {Chang}\ \emph {et~al.}(2013)\citenamefont {Chang},
  \citenamefont {H\"{o}lzer}, \citenamefont {Travers},\ and\ \citenamefont
  {{P.~{St.J}.~Russell}}}]{Chang13}%
  \BibitemOpen
  \bibfield  {author} {\bibinfo {author} {\bibfnamefont {W.}~\bibnamefont
  {Chang}}, \bibinfo {author} {\bibfnamefont {P.}~\bibnamefont {H\"{o}lzer}},
  \bibinfo {author} {\bibfnamefont {J.~C.}\ \bibnamefont {Travers}}, \ and\
  \bibinfo {author} {\bibnamefont {{P.~{St.J}.~Russell}}},\ }\href@noop {}
  {\bibfield  {journal} {\bibinfo  {journal} {Opt. Lett}\ }\textbf {\bibinfo
  {volume} {38}},\ \bibinfo {pages} {2984} (\bibinfo {year}
  {2013})}\BibitemShut {NoStop}%
\bibitem [{\citenamefont {Saleh}\ \emph {et~al.}(2012)\citenamefont {Saleh},
  \citenamefont {Chang}, \citenamefont {Travers}, \citenamefont
  {{P.~{St.J}.~Russell}},\ and\ \citenamefont {Biancalana}}]{Saleh12}%
  \BibitemOpen
  \bibfield  {author} {\bibinfo {author} {\bibfnamefont {M.~F.}\ \bibnamefont
  {Saleh}}, \bibinfo {author} {\bibfnamefont {W.}~\bibnamefont {Chang}},
  \bibinfo {author} {\bibfnamefont {J.~C.}\ \bibnamefont {Travers}}, \bibinfo
  {author} {\bibnamefont {{P.~{St.J}.~Russell}}}, \ and\ \bibinfo {author}
  {\bibfnamefont {F.}~\bibnamefont {Biancalana}},\ }\href@noop {} {\bibfield
  {journal} {\bibinfo  {journal} {Phys. Rev. Lett.}\ }\textbf {\bibinfo
  {volume} {109}},\ \bibinfo {pages} {113902} (\bibinfo {year}
  {2012})}\BibitemShut {NoStop}%
\bibitem [{\citenamefont {Saleh}\ \emph {et~al.}(2014)\citenamefont {Saleh},
  \citenamefont {Marini},\ and\ \citenamefont {Biancalana}}]{Saleh14}%
  \BibitemOpen
  \bibfield  {author} {\bibinfo {author} {\bibfnamefont {M.~F.}\ \bibnamefont
  {Saleh}}, \bibinfo {author} {\bibfnamefont {A.}~\bibnamefont {Marini}}, \
  and\ \bibinfo {author} {\bibfnamefont {F.}~\bibnamefont {Biancalana}},\
  }\href@noop {} {\bibfield  {journal} {\bibinfo  {journal} {Phys. Rev. A}\
  }\textbf {\bibinfo {volume} {89}},\ \bibinfo {pages} {023801} (\bibinfo
  {year} {2014})}\BibitemShut {NoStop}%
\bibitem [{\citenamefont {Yoshikawa}\ and\ \citenamefont
  {Imasaka}(1993)}]{Yoshikawa93}%
  \BibitemOpen
  \bibfield  {author} {\bibinfo {author} {\bibfnamefont {S.}~\bibnamefont
  {Yoshikawa}}\ and\ \bibinfo {author} {\bibfnamefont {T.}~\bibnamefont
  {Imasaka}},\ }\href@noop {} {\bibfield  {journal} {\bibinfo  {journal} {Opt.
  Comm.}\ }\textbf {\bibinfo {volume} {96}},\ \bibinfo {pages} {94} (\bibinfo
  {year} {1993})}\BibitemShut {NoStop}%
\bibitem [{\citenamefont {Kaplan}(1994)}]{Kaplan94}%
  \BibitemOpen
  \bibfield  {author} {\bibinfo {author} {\bibfnamefont {A.~E.}\ \bibnamefont
  {Kaplan}},\ }\href@noop {} {\bibfield  {journal} {\bibinfo  {journal} {Phys.
  Rev. Lett.}\ }\textbf {\bibinfo {volume} {73}},\ \bibinfo {pages} {1243�}
  (\bibinfo {year} {1994})}\BibitemShut {NoStop}%
\bibitem [{\citenamefont {Kawano}\ \emph {et~al.}(1998)\citenamefont {Kawano},
  \citenamefont {Hirakawa},\ and\ \citenamefont {Imasaka}}]{Kawano98}%
  \BibitemOpen
  \bibfield  {author} {\bibinfo {author} {\bibfnamefont {H.}~\bibnamefont
  {Kawano}}, \bibinfo {author} {\bibfnamefont {Y.}~\bibnamefont {Hirakawa}}, \
  and\ \bibinfo {author} {\bibfnamefont {T.}~\bibnamefont {Imasaka}},\
  }\href@noop {} {\bibfield  {journal} {\bibinfo  {journal} {IEEE. J. Quantum
  Electron.}\ }\textbf {\bibinfo {volume} {34}},\ \bibinfo {pages} {260}
  (\bibinfo {year} {1998})}\BibitemShut {NoStop}%
\bibitem [{\citenamefont {Nazarkin}\ \emph {et~al.}(1999)\citenamefont
  {Nazarkin}, \citenamefont {Korn}, \citenamefont {Wittmann},\ and\
  \citenamefont {Elsaesser}}]{Nazarkin99}%
  \BibitemOpen
  \bibfield  {author} {\bibinfo {author} {\bibfnamefont {A.}~\bibnamefont
  {Nazarkin}}, \bibinfo {author} {\bibfnamefont {G.}~\bibnamefont {Korn}},
  \bibinfo {author} {\bibfnamefont {M.}~\bibnamefont {Wittmann}}, \ and\
  \bibinfo {author} {\bibfnamefont {T.}~\bibnamefont {Elsaesser}},\ }\href@noop
  {} {\bibfield  {journal} {\bibinfo  {journal} {Phys. Rev. Lett.}\ }\textbf
  {\bibinfo {volume} {83}},\ \bibinfo {pages} {2560} (\bibinfo {year}
  {1999})}\BibitemShut {NoStop}%
\bibitem [{\citenamefont {Kalosha}\ and\ \citenamefont
  {Herrmann}(2000)}]{Kalosha00}%
  \BibitemOpen
  \bibfield  {author} {\bibinfo {author} {\bibfnamefont {V.~P.}\ \bibnamefont
  {Kalosha}}\ and\ \bibinfo {author} {\bibfnamefont {J.}~\bibnamefont
  {Herrmann}},\ }\href@noop {} {\bibfield  {journal} {\bibinfo  {journal}
  {Phys. Rev. Lett.}\ }\textbf {\bibinfo {volume} {85}},\ \bibinfo {pages}
  {1226} (\bibinfo {year} {2000})}\BibitemShut {NoStop}%
\bibitem [{\citenamefont {Korn}\ \emph {et~al.}(1998)\citenamefont {Korn},
  \citenamefont {D\"{u}hr},\ and\ \citenamefont {Nazarkin}}]{Korn98}%
  \BibitemOpen
  \bibfield  {author} {\bibinfo {author} {\bibfnamefont {G.}~\bibnamefont
  {Korn}}, \bibinfo {author} {\bibfnamefont {O.}~\bibnamefont {D\"{u}hr}}, \
  and\ \bibinfo {author} {\bibfnamefont {A.}~\bibnamefont {Nazarkin}},\
  }\href@noop {} {\bibfield  {journal} {\bibinfo  {journal} {Phys. Rev. Lett.}\
  }\textbf {\bibinfo {volume} {81}},\ \bibinfo {pages} {1215} (\bibinfo {year}
  {1998})}\BibitemShut {NoStop}%
\bibitem [{\citenamefont {Ihara}\ \emph {et~al.}(2006)\citenamefont {Ihara},
  \citenamefont {Eshima}, \citenamefont {Zaitsu}, \citenamefont {Kamitomo},
  \citenamefont {Shinzen}, \citenamefont {Hirakawa},\ and\ \citenamefont
  {Imasaka}}]{Ihara06}%
  \BibitemOpen
  \bibfield  {author} {\bibinfo {author} {\bibfnamefont {K.}~\bibnamefont
  {Ihara}}, \bibinfo {author} {\bibfnamefont {C.}~\bibnamefont {Eshima}},
  \bibinfo {author} {\bibfnamefont {S.}~\bibnamefont {Zaitsu}}, \bibinfo
  {author} {\bibfnamefont {S.}~\bibnamefont {Kamitomo}}, \bibinfo {author}
  {\bibfnamefont {K.}~\bibnamefont {Shinzen}}, \bibinfo {author} {\bibfnamefont
  {Y.}~\bibnamefont {Hirakawa}}, \ and\ \bibinfo {author} {\bibfnamefont
  {T.}~\bibnamefont {Imasaka}},\ }\href@noop {} {\bibfield  {journal} {\bibinfo
   {journal} {Appl. Phys. Lett.}\ }\textbf {\bibinfo {volume} {88}},\ \bibinfo
  {pages} {074101} (\bibinfo {year} {2006})}\BibitemShut {NoStop}%
\bibitem [{\citenamefont {Belli}\ \emph {et~al.}(2014)\citenamefont {Belli},
  \citenamefont {Abdolvand}, \citenamefont {Chang}, \citenamefont {Travers},\
  and\ \citenamefont {{P.~{St.J}.~Russell}}}]{Belli14}%
  \BibitemOpen
  \bibfield  {author} {\bibinfo {author} {\bibfnamefont {F.}~\bibnamefont
  {Belli}}, \bibinfo {author} {\bibfnamefont {A.}~\bibnamefont {Abdolvand}},
  \bibinfo {author} {\bibfnamefont {W.}~\bibnamefont {Chang}}, \bibinfo
  {author} {\bibfnamefont {J.~C.}\ \bibnamefont {Travers}}, \ and\ \bibinfo
  {author} {\bibnamefont {{P.~{St.J}.~Russell}}},\ }\href@noop {} {\bibfield
  {journal} {\bibinfo  {journal} {in CLEO: 2014, OSA Technical Digest (online)
  (Optical Society of America, 2014), paper FW1D.1}\ } (\bibinfo {year}
  {2014})}\BibitemShut {NoStop}%
\bibitem [{\citenamefont {Wittmann}\ \emph {et~al.}(2000)\citenamefont
  {Wittmann}, \citenamefont {Nazarkin},\ and\ \citenamefont
  {Korn}}]{Wittmann00}%
  \BibitemOpen
  \bibfield  {author} {\bibinfo {author} {\bibfnamefont {M.}~\bibnamefont
  {Wittmann}}, \bibinfo {author} {\bibfnamefont {A.}~\bibnamefont {Nazarkin}},
  \ and\ \bibinfo {author} {\bibfnamefont {G.}~\bibnamefont {Korn}},\
  }\href@noop {} {\bibfield  {journal} {\bibinfo  {journal} {Phys. Rev. Lett.}\
  }\textbf {\bibinfo {volume} {84}},\ \bibinfo {pages} {5508} (\bibinfo {year}
  {2000})}\BibitemShut {NoStop}%
\bibitem [{\citenamefont {Wannier}(1960)}]{Wannier60}%
  \BibitemOpen
  \bibfield  {author} {\bibinfo {author} {\bibfnamefont {G.~H.}\ \bibnamefont
  {Wannier}},\ }\href@noop {} {\bibfield  {journal} {\bibinfo  {journal} {Phys.
  Rev.}\ }\textbf {\bibinfo {volume} {117}},\ \bibinfo {pages} {432} (\bibinfo
  {year} {1960})}\BibitemShut {NoStop}%
\bibitem [{\citenamefont {Bloch}(1928)}]{Bloch28}%
  \BibitemOpen
  \bibfield  {author} {\bibinfo {author} {\bibfnamefont {F.}~\bibnamefont
  {Bloch}},\ }\href@noop {} {\bibfield  {journal} {\bibinfo  {journal} {Z.
  Phys.}\ }\textbf {\bibinfo {volume} {52}},\ \bibinfo {pages} {555} (\bibinfo
  {year} {1928})}\BibitemShut {NoStop}%
\bibitem [{\citenamefont {Zener}(1934)}]{Zener34}%
  \BibitemOpen
  \bibfield  {author} {\bibinfo {author} {\bibfnamefont {C.}~\bibnamefont
  {Zener}},\ }\href@noop {} {\bibfield  {journal} {\bibinfo  {journal} {R. Soc.
  Lond. A}\ }\textbf {\bibinfo {volume} {145}},\ \bibinfo {pages} {523}
  (\bibinfo {year} {1934})}\BibitemShut {NoStop}%
\bibitem [{\citenamefont {Butylkin}\ \emph {et~al.}(1989)\citenamefont
  {Butylkin}, \citenamefont {Kaplan}, \citenamefont {Khronopulo},\ and\
  \citenamefont {Yakubovich}}]{Butylkin89}%
  \BibitemOpen
  \bibfield  {author} {\bibinfo {author} {\bibfnamefont {V.~S.}\ \bibnamefont
  {Butylkin}}, \bibinfo {author} {\bibfnamefont {A.~E.}\ \bibnamefont
  {Kaplan}}, \bibinfo {author} {\bibfnamefont {Y.~G.}\ \bibnamefont
  {Khronopulo}}, \ and\ \bibinfo {author} {\bibfnamefont {E.~I.}\ \bibnamefont
  {Yakubovich}},\ }\href@noop {} {\emph {\bibinfo {title} {Resonant Nonlinear
  Interaction of Light with Matter}}},\ \bibinfo {edition} {1st}\ ed.\
  (\bibinfo  {publisher} {Springer-Verlag Berlin Heidelberg},\ \bibinfo {year}
  {1989})\BibitemShut {NoStop}%
\bibitem [{\citenamefont {Agrawal}(2007)}]{Agrawal07}%
  \BibitemOpen
  \bibfield  {author} {\bibinfo {author} {\bibfnamefont {G.~P.}\ \bibnamefont
  {Agrawal}},\ }\href@noop {} {\emph {\bibinfo {title} {Nonlinear Fiber
  Optics}}},\ \bibinfo {edition} {4th}\ ed.,\ San Diego, California\ (\bibinfo
  {publisher} {Academic Press},\ \bibinfo {year} {2007})\BibitemShut {NoStop}%
\bibitem [{\citenamefont {Gagnon}\ and\ \citenamefont
  {B\'{e}langer}(1990)}]{Gagnon90}%
  \BibitemOpen
  \bibfield  {author} {\bibinfo {author} {\bibfnamefont {L.}~\bibnamefont
  {Gagnon}}\ and\ \bibinfo {author} {\bibfnamefont {P.~A.}\ \bibnamefont
  {B\'{e}langer}},\ }\href@noop {} {\bibfield  {journal} {\bibinfo  {journal}
  {Opt. Lett}\ }\textbf {\bibinfo {volume} {15}},\ \bibinfo {pages} {466}
  (\bibinfo {year} {1990})}\BibitemShut {NoStop}%
\bibitem [{\citenamefont {Bouchard}\ and\ \citenamefont
  {Luban}(1995)}]{Bouchard95}%
  \BibitemOpen
  \bibfield  {author} {\bibinfo {author} {\bibfnamefont {A.~M.}\ \bibnamefont
  {Bouchard}}\ and\ \bibinfo {author} {\bibfnamefont {M.}~\bibnamefont
  {Luban}},\ }\href@noop {} {\bibfield  {journal} {\bibinfo  {journal} {Phys.
  Rev. B}\ }\textbf {\bibinfo {volume} {52}},\ \bibinfo {pages} {5105}
  (\bibinfo {year} {1995})}\BibitemShut {NoStop}%
\bibitem [{\citenamefont {Iafrate}\ \emph {et~al.}(1998)\citenamefont
  {Iafrate}, \citenamefont {Reynolds}, \citenamefont {He}, ,\ and\
  \citenamefont {Krieger}}]{Iafrate98}%
  \BibitemOpen
  \bibfield  {author} {\bibinfo {author} {\bibfnamefont {G.~J.}\ \bibnamefont
  {Iafrate}}, \bibinfo {author} {\bibfnamefont {J.~P.}\ \bibnamefont
  {Reynolds}}, \bibinfo {author} {\bibfnamefont {J.}~\bibnamefont {He}}, \
  and\ \bibinfo {author} {\bibfnamefont {J.~B.}\ \bibnamefont {Krieger}},\
  }\href@noop {} {\bibfield  {journal} {\bibinfo  {journal} {Int. Journal of
  High Speed Electr.}\ }\textbf {\bibinfo {volume} {9}},\ \bibinfo {pages}
  {223} (\bibinfo {year} {1998})}\BibitemShut {NoStop}%
\end{thebibliography}
\end{document}